\documentclass[pra,nobalancelastpage,twocolumn,nofootinbib,superscriptaddress]{revtex4-2}
\pdfoutput=1
\usepackage[utf8]{inputenc}
\usepackage{color,amsthm,amsmath,amsxtra,amsfonts,dsfont,graphicx,bm,amssymb}
\usepackage{qcircuit}
\usepackage{booktabs}
\usepackage{verbatim}
\usepackage{hyperref}
\usepackage[english]{babel}
\usepackage[dvipsnames]{xcolor}
\usepackage{mathtools}
\usepackage{tikz}
\usetikzlibrary{shapes, arrows, positioning, calc}
\usepackage{braket}

\newcommand{\tr}[1]{{\mathrm{Tr}\left[#1\right]}}

\newcommand{\di}{\mathrm{d}}
\begin{document}
\title{Quantum Simulation of Dynamical Response Functions of Equilibrium States}
\author{Esther Cruz}
\author{Dominik S.~Wild}
\author{Mari Carmen Ba\~nuls}
\author{J.~Ignacio Cirac}
\affiliation{Max-Planck-Institut für Quantenoptik, Hans-Kopfermann-Str.~1, D-85748 Garching, Germany}
\affiliation{Munich Center for Quantum Science and Technology (MCQST), Schellingstr.~4, D-80799 München, Germany}

\begin{abstract}
   The computation of dynamical response functions is central to many problems in condensed matter physics. Owing to the rapid growth of quantum correlations following a quench, classical methods face significant challenges even if an efficient description of the equilibrium state is available. Quantum computing offers a promising alternative. However, existing approaches often assume access to the equilibrium state, which may be difficult to prepare in practice. In this work, we present a method that circumvents this by using energy filter techniques, enabling the computation of response functions and other dynamical properties in both microcanonical and canonical ensembles. Our approach only requires the preparation of states that have significant weight at the desired energy. The dynamical response functions are then reconstructed from measurements after quenches of varying duration by classical postprocessing. We illustrate the algorithm numerically by applying it to compute the dynamical conductivity of a free-fermion model, which unveils the energy-dependent localization properties of the model.
\end{abstract}
\maketitle

\paragraph*{Introduction.}
A key objective of condensed matter physics is to uncover the properties of equilibrium phases of matter. A great variety of classical computational methods has been developed to this end. For instance, tensor networks~\cite{cirac_matrix_2021} efficiently describe one-dimensional systems with low entanglement, which renders them an ideal tool for the study of ground states of local, gapped Hamiltonians. In higher dimensions and at finite temperature, quantum Monte Carlo methods~\cite{Acioli1997,Foulkes2001,Sandvik2010} are a powerful alternative approach for Hamiltonians without a sign problem. However, neither of these methods is well suited for computing dynamical response functions such as the frequency-dependent electrical conductivity or the magnetic susceptibility because time-dependent perturbations away from equilibrium typically lead to a rapid growth of entanglement and introduce a complex sign problem~\cite{cohen_taming_2015}.

\begin{figure*}
    
\includegraphics[width=\linewidth]{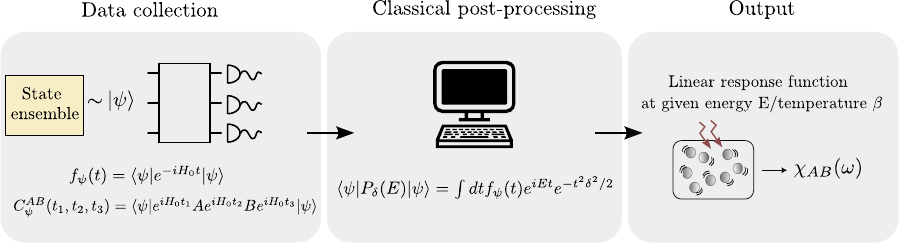}
\caption{Scheme of the algorithm. The first step is the data collection. A state $\ket{\psi}$ is drawn from an ensemble of easy-to-prepare states. From a sequence of projective measurements, one can estimate the functions $f_\psi(t)$ and $C_\psi^{AB}(t_1,t_2,t_3)$ for a series of times $t,t_1,t_2,t_3$. The real-time functions are then fed into a classical post-processing step. This consists of performing discrete approximations to Fourier transforms to recover the local density of states and the response functions evaluated on the filtered state, both in the time and frequency domain. In this step, one may also compute the average of such functions over many instances of initial states $\ket{\psi}$. The final output is the response function $\chi_{AB}(\omega)$ for the desired range of frequencies at a given target energy or temperature.}
\label{fig:scheme-algorithm}
\end{figure*}

Quantum computers offer a promising avenue to overcoming these limitations as they enable the efficient simulation of quantum many-body dynamics. Several recent studies have explored different methods of measuring dynamical response functions on a quantum computer~\cite{roggero_linear_2019, endo_calculation_2020, libbi_effective_2022,kokcu_linear_2023, maskara_programmable_2025, loaiza_nonlinear_2024}.  However, these works assume that the equilibrium state of interest is provided as an input to the computation. In practice, such a state may be difficult to prepare and may require significant quantum resources. In this work, we address this challenge by combining the measurement of dynamical response functions with spectral filters~\cite{ge_faster_2019,somma_quantum_2019,lu_algorithms_2021,lin_heisenberg_2022,Morettini_2024, Hemery_2024}, as summarized in Fig.~\ref{fig:scheme-algorithm}. Our method proceeds by measuring the expectation value of products of observables and the time-evolution unitary at different times. By classically post-processing the collected data, we are able to project the expectation values onto a narrow energy range and can thus infer dynamical response functions of equilibrium states.

Instead of having to prepare an equilibrium state, it suffices for our method to prepare states with sufficient overlap with the energy range of interest. The total runtime required to estimate an expectation value of a state $\ket{\psi}$ projected onto an energy range of width $\delta$ around $E$ is polynomial in $1/r_\delta(E)$, where $r_\delta(E)$ quantifies the overlap of $\ket{\psi}$ with the energy eigenstates in this window. The maximum depth of the quantum circuits to be executed depends polynomially on $1/\delta$~\cite{lu_algorithms_2021}.
The choice of initial state determines the relative contribution from each energy eigenstate within the energy range. Nevertheless, in many physical scenarios, the filtered observable becomes independent of the detailed properties of the initial state for small values of $\delta$. This follows, for instance, from the eigenstate thermalization hypothesis (ETH)~\cite{deutsch_quantum_1991, srednicki_chaos_1994,rigol_thermalization_2008}, according to which local observables of energy eigenstates are smooth functions of the energy.  
The dependence on the initial state may also be reduced by averaging over a set of states, which moreover provides access to the canonical ensemble when weighting the contributions at different energies with the Boltzmann factor~\cite{lu_algorithms_2021}.
 
Our method offers a general-purpose framework for computing dynamical response functions of equilibrium states. 
The range of energies and temperatures that can be probed using our approach depends on the model of interest and on the given state preparation capabilities. We cannot expect to be able to reach arbitrarily low temperatures for general, local Hamiltonians, since the determination of the ground state energy is a QMA-complete problem~\cite{kempe_complexity_2005}. However, quantum phases corresponding to this worst-case hardness are unlikely to occur in nature. We therefore expect that for a wide variety of problems in physics, the required initial states can be efficiently prepared on a quantum computer.

\paragraph*{Dynamical response functions.}
Before describing our algorithm in detail and applying it to a concrete model, we review the main concepts of dynamical response functions. We consider a Hamiltonian of the form $H(t) = H_0 + g(t) B$. Here, $H_0$ governs the equilibrium properties while $g(t)$ is the time-dependent amplitude of a perturbation that couples to the operator $B$. The effect of the perturbation can be probed by the change in the expectation value of some relevant observable $A$. To the lowest order in perturbation theory, this change $\delta A(t) := \tr{e^{-iHt} A e^{iHt} \rho_0} - \tr{ e^{-iH_0 t} A e^{iH_0 t} \rho_0}$ is given by the  Kubo formula~\cite{kubo_statistical-mechanical_1957}:
\begin{equation}
    \delta A(t) =  - i \int_0^t \di t' g(t') \, \mathrm{Tr} \left( [A(t),B(t')] \rho_0 \right) ,
    \label{eq:kubo}
\end{equation}
where  $\rho_0$ is the state of the system at time $t = 0$. In Eq.~\eqref{eq:kubo}, the time-dependent operators are understood to evolve in the interaction picture, i.e., $A(t) = e^{i H_0 t} A e^{- i H_0 t}$ and similarly for $B(t')$.

While the above formula holds for any state, we are typically interested in equilibrium states with respect to the unperturbed Hamiltonian, such as the ground state or Gibbs states at finite temperature. In this case, the initial state $\rho_0$ commutes with $H_0$, such that the commutator $\chi_{AB}(t-t') = -i \, \mathrm{Tr} \left( [A(t), B(t')] \rho_0 \right)$, also referred to as the linear response function, only depends on the time difference $t - t'$. By introducing the Fourier transform 
\begin{align} 
    \chi_{AB}(\omega) = \int_{0}^{\infty} \di t \, e^{i\omega t} \chi_{AB} (t) \ ,
    \label{eq:correlation-frequency}
\end{align}
the Kubo formula can be brought into the convenient form $\delta A(\omega) = g(\omega) \chi_{AB}(\omega)$.

\paragraph*{Energy filters.}
A naive approach to measuring linear response functions on a quantum computer proceeds by directly preparing the equilibrium state $\rho_0$, which may, however, be a challenging task. As an alternative, recent works \cite{somma_quantum_2019,lu_algorithms_2021, schuckert_probing_2023, morettini_energy-filtered_2024} have proposed to recover expectation values of equilibrium states by classically postprocessing time series data obtained from measurements of states that are easier to prepare in practice. The approach is based on the observation that an operator of the form
\begin{equation}
    P_\delta(E) = \frac{1}{\sqrt{2 \pi \delta^2}} e^{-(H_0-E)^2/2 \delta^2} 
\end{equation}
may be expressed in terms of its Fourier transform
\begin{equation}
    P_\delta(E) = \int \frac{\di t}{2 \pi} \, e^{-i (H_0 - E) t} e^{- t^2 \delta^2 / 2}  ,
    \label{eq:filter_fourier}
\end{equation}
corresponding to a linear combination of the time evolution operator $e^{- i H_0 t}$ at different times $t$. The operator $P_\delta(E)$ suppresses the amplitudes of energy eigenstates with eigenvalues far from $E$ and may thus be viewed as a filter of width $\delta$. We note that this approach not only applies to the Gaussian filter considered here but also to other choices of filter functions~\cite{lu_algorithms_2021}.

In practice, we may wish to apply the filter to a state $\ket{\psi}$ that can be efficiently prepared on a quantum computer. The expectation value of an observable $A$ in the filtered state ${P_\delta(E) \ket{\psi}} / {\Vert P_\delta(E) \ket{\psi} \Vert }$ can be written as
\begin{equation} \label{eq:exp-value-filtered-state}
  A_\psi(\delta,E) =   \frac{\bra{\psi} P_\delta(E) AP_\delta(E) \ket{\psi} }{\bra{\psi} P^2_\delta(E) \ket{\psi}} .
\end{equation}
Using Eq.~\eqref{eq:filter_fourier}, this value can be computed from measurements of $\braket{\psi | e^{i H_0 t} A e^{- i H_0 t'} | \psi}$ and $\braket{\psi | e^{- i H_0 t} | \psi}$. We emphasize that the filtered state is never directly prepared, but the expectation value is inferred from independent measurements of time-dependent quantities. This procedure is efficient provided the denominator in Eq.~\eqref{eq:exp-value-filtered-state} is sufficiently large~\cite{lu_algorithms_2021}, meaning that the initial state $\ket{\psi}$ has sufficient overlap with the energy window of width $\delta$ around $E$.

In the limit $\delta \to 0$, the expectation value  $A_\psi(\delta, E)$ converges to that of the energy eigenvector with eigenvalue closest to $E$. This limit is, however, not feasible as it would require a value of $\delta$ that is exponentially small in the system size, leading to exponentially large evolution times. Fortunately, in many physical systems, local observables equilibrate rapidly such that much larger values of $\delta$ suffice. 

Instead of applying the filter to an individual pure state, we may also consider the filter itself as a mixed state, $\rho_\delta(E) = {P_\delta(E)} / {\tr{P_\delta(E)}}$. We will refer to this state as the filter ensemble. It converges to the microcanonical ensemble in the limit $\delta \to 0$~\cite{yang_classical_2022}. Using a complete set of states $\{ \ket{\psi} \}$, the expectation value of an observable $A$ in the filter ensemble can be written as
\begin{align}
   \frac{\tr{A P_\delta (E) }}{\tr{P_\delta(E)}}  & = \sum_\psi p_\psi {A}_\psi(\delta / \sqrt{2}, E) \ ,  \label{eq:exp-value-filter-ensemble}
\end{align}
where $p_\psi = \braket{\psi | P_\delta(E) | \psi} / \sum_{\psi'} \braket{\psi' | P_\delta(E) | \psi'}$ defines a probability distribution over the set of states. The expectation value in Eq.~\eqref{eq:exp-value-filter-ensemble} can thus be estimated empirically by computing $A_\psi(\delta/\sqrt{2}, E)$ for random states, where each state is drawn with a probability proportional to $\braket{\psi | P_\delta(E) | \psi}$, corresponding to the overlap of the state with the energy window of interest. Alternatively, Eq.~\eqref{eq:exp-value-filter-ensemble} can be written in terms of the quantity $\tilde{A}_\psi(\delta, E) = \braket{\psi | A P_\delta(E) | \psi} / \braket{\psi | P_\delta(E) | \psi}$ instead of $A_\psi(\delta / \sqrt{2}, E)$. This has the advantage that it is sufficient to measure $\braket{ \psi | A e^{-i H_0 t} | \psi}$, which depends on a single time variable instead of the two time variables in $\braket{ \psi | e^{-i H_0 t} A e^{-i H_0 t'} | \psi}$. This approach to measuring properties of the microcanonical ensemble can also be extended to the canonical ensemble by averaging over the energy $E$, weighted by the Boltzmann factor~\cite{lu_algorithms_2021}.

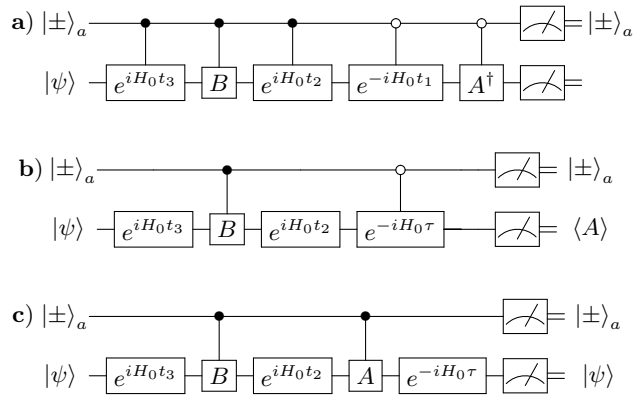
\begin{figure}[t]
    \centering
  \centering
    \[
    \Qcircuit @C=0.7em @R=1em {
        \mathbf{a)} \qquad \qquad  & \ket{\pm}_a   \qquad  & \ctrl{1} & \ctrl{1}  &  \ctrl{1}  & \ctrlo{1}  &  \ctrlo{1}   & \meter & \cw &  \quad \ket{\pm}_a \\
      \phantom{a) \qquad \qquad} &   \lstick{\ket{\psi}}  &  \gate{e^{iH_0 t_3}} & \gate{B} & \gate{e^{iH_0 t_2}} & \gate{e^{-iH_0 t_1}} &  \gate{A^\dagger} &  \meter & \cw  \\
    }
    \]
    
    \[
    \Qcircuit @C=0.7em @R=1em {
         \mathbf{b)}  \qquad  \qquad &  \ket{\pm}_a   \qquad     & \qw  & \ctrl{1}  &  \qw  & \ctrlo{1} & \qw  & \qw & \meter & \cw   & & \ket{\pm}_a \\
     \phantom{b) \qquad \qquad} & \lstick{\ket{\psi}} &    \gate{e^{iH_0 t_3}} & \gate{B} & \gate{e^{iH_0 t_2}} & \gate{e^{-iH_0 \tau}} &  \phantom{\gate{A^\dagger}}& \qw & \meter & \cw & & \langle A \rangle  \\
    }
    \]
    
    \[
    \Qcircuit @C=0.7em @R=1em {
       \mathbf{c)} \qquad \qquad & \ket{\pm}_a   \qquad    & \qw      & \ctrl{1}  &  \qw &   \ctrl{1}   & \qw & \meter & \cw  & & \ket{ \pm}_a\\
     \phantom{c) \qquad \qquad} & \lstick{\ket{\psi}}   &  \gate{e^{iH_0 t_3}} & \gate{B} & \gate{e^{iH_0 t_2}} &  \gate{A} & \gate{e^{-iH_0 \tau}} & \meter & \cw & & \ket{\psi } \\
    }
    \]
    \caption{Quantum circuits to measure the quantity in Eq.~\eqref{eq:quantities-times}. The first register is a single-qubit ancilla, and $\ket{\pm}_a$ indicates the basis vector in the $a \in \{x,y \}$ basis. Circuit a) is a generalization of the Hadamard test~\cite{somma_simulating_2002}. This circuit can be simplified by eliminating some of the controlled evolutions and measuring the observable $A$, as shown in b), where  $\tau = t_1 + t_2 + t_3$. In c), we provide an alternative way of measuring \eqref{eq:quantities-times} that does not require controlling any evolution operators but only the observables $A$ and $B$.}
    \label{fig:quantum_circuits}
\end{figure}

\paragraph*{Quantum simulation of response functions.}
In this work, we apply the above filter formalism to dynamical response functions. For a sufficiently small filter width $\delta$, filtered pure states are approximately stationary, such that we assume that $\braket{\psi| P_\delta(E) [ A(t), B(t')] P_\delta(E) |\psi}$ depends only on $t - t'$. Note that this holds exactly for $\mathrm{Tr} \{ [A(t), B(t')] P_\delta(E) \}$ since the filter ensemble is stationary. Following the above reasoning for a single observable $A$, we find that $\chi^{AB}(t - t')$ can be obtained by measuring the quantities
\begin{align} 
    f_{\psi} (t) &= \bra{\psi} e^{-iH_0 t} \ket{\psi} \label{eq:loschmidt} , \\
    C^{AB}_\psi (t_1,t_2,t_3) & = \langle \psi | e^{iH_0 t_1}A e^{iH_0 t_2} B e^{iH_0 t_3} \ket{\psi} .\label{eq:quantities-times}
\end{align}
We highlight that it is necessary to measure the magnitude and phase of both quantities. For the first quantity, known as the Loschmidt echo, this can be done by applying a Hadamard test or using recently developed approaches that avoid global control of the time-evolution unitary~\cite{lu_algorithms_2021, yang_phase-sensitive_2024, clinton_quantum_2024, maskara_programmable_2025, chan_algorithmic_2025}. The second quantity can also be measured using an extended version of the Hadamard test, shown in Fig.~\ref{fig:quantum_circuits}(a), provided the operators $A$ and $B$ are unitary. This assumption is not restrictive as any observable may be decomposed into a sum of Pauli operators. 

As the application of consecutive Hadamard tests may be cumbersome in practice, we may wish to simplify the circuit in Fig.~\ref{fig:quantum_circuits}(a). First, we observe that the control of the first time-evolution unitary can be removed if we adjust the time in the last unitary from $t_1$ to $t_1 - t_3$. If the observable $A$ can be directly measured, then it is possible to remove further gates as shown in Fig.~\ref{fig:quantum_circuits}(b). Alternatively, one can perform a projective measurement onto the initial state. If the initial state was prepared by applying a unitary to a computational basis state, such a measurement can be implemented by reversing the state preparation and measuring in the computational basis.   When applied to the circuit shown in Fig.~\ref{fig:quantum_circuits}(c), this measurement yields the quantity $\bra{\psi} e^{iH t_1} A e^{iH t_2} B e^{iH t_3}  \ket{\psi} \bra{\psi} e^{-iH (t_1+t_2+t_3)} \ket{\psi}$. By dividing by the separately measured Loschmidt echo $ \bra{\psi} e^{-iH (t_1+t_2+t_3)} \ket{\psi}$, we thus obtain $C_\psi^{AB}(t_1, t_2, t_3)$. We highlight that the circuit in Fig.~\ref{fig:quantum_circuits}(c) only requires locally controlled gates if $A$ and $B$ are local operators.

In practice, it is only possible to measure $f_\psi(t)$ and $C_\psi^{AB}(t_1, t_2, t_3)$ for a discrete set of times. For this reason, it is necessary to discretize the integral in Eq.~\eqref{eq:filter_fourier}. Here, we will choose a simple discretization in terms of the Riemann sum
\begin{align}
    P_\delta(E) \approx \sum_{k = -K}^{K} \Delta t \, e^{-i (H - E) \Delta t \, k} e^{- (\delta \Delta t)^2 k^2 / 2} ,
   \label{eq:filter-sum}
\end{align}
although other approximations or intrinsically discrete filters such as the cosine filter may be used instead~\cite{lu_algorithms_2021}. The time step $\Delta t$ should be chosen such that the maximum frequency present in the system can be resolved. For local Hamiltonians with bounded interactions, it thus suffices to choose $\Delta t$ inversely proportional to the system size since the energy bandwidth increases at most linearly with it. The cutoff $K$ depends on the filter width and should be chosen such that $K \delta \Delta t \gg 1$, in which case truncation error is suppressed by $\exp[- O(K^2 \delta^2 \Delta t^2)]$.

\label{section:illustration-algorithm}
\begin{figure*}
    \centering
    \includegraphics[width=\linewidth]{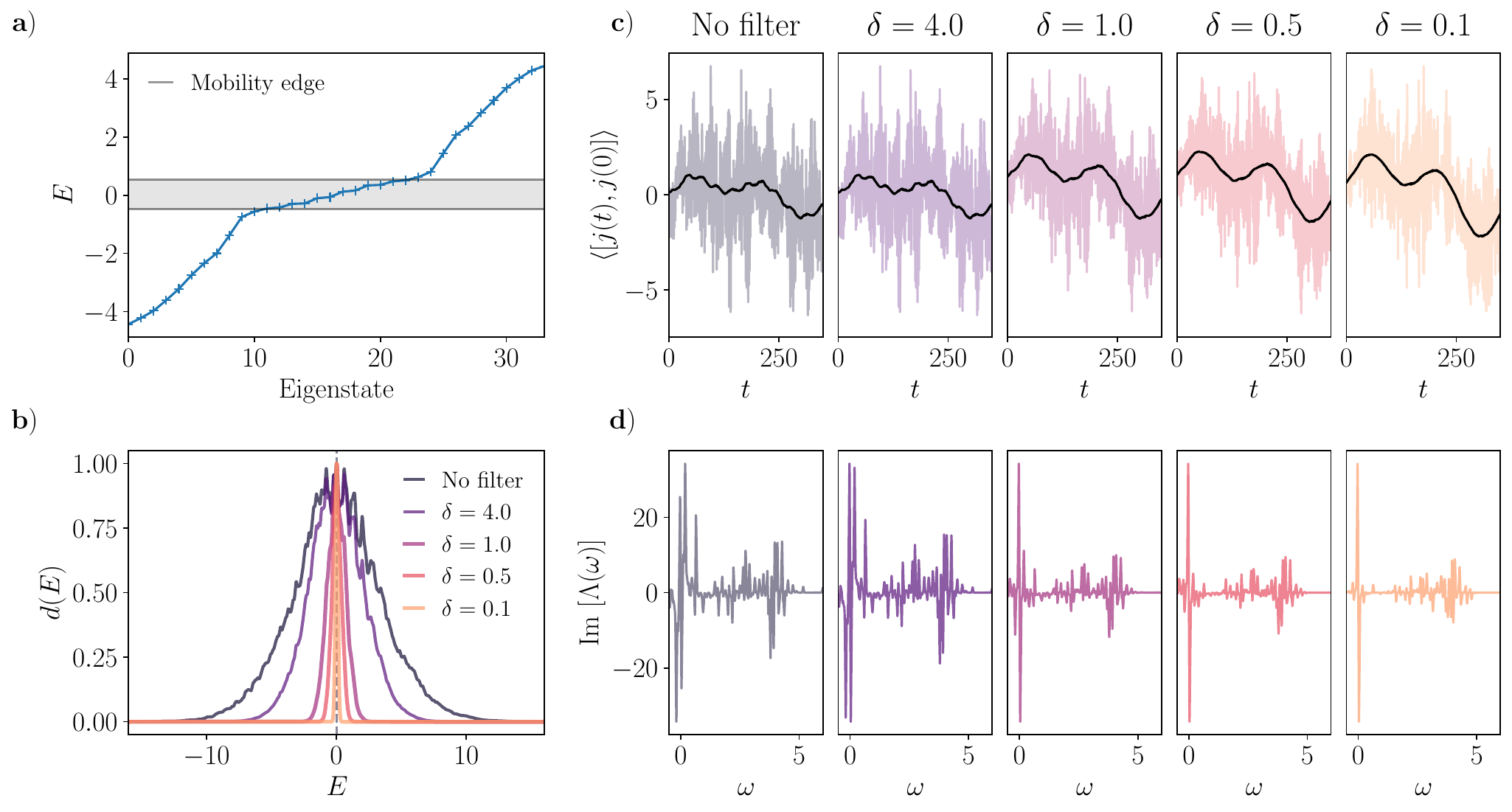}
    \caption{ a) Single-particle spectrum of the Hamiltonian. The eigenstates in the gray shaded region in the middle of the spectrum are extended, whereas all other eigenstates are localized.  b) Local density of states of $\ket{\Psi}$ and $ P_\delta(\bar{E}) \ket{\Psi} / \Vert P_\delta(\bar{E}) \ket{\Psi} \Vert$ for different values of the filter width $\delta$. Here, $\ket{\Psi}$ is a Fock state of $N_0 = 6$ fermions in a system comprising $N = 34$ sites. The fermions are randomly distributed across the odd sites such that $\bar{E} = \braket{\Psi | H | \Psi} = 0$. The LDOS $d(E)$ is computed through an energy filter with a small width $\eta = 0.05$. The maximum value of each curve is scaled to $1$ for visibility.  c)~Expectation value of $[j(t),j(0)]$ for the same states as in b). The black solid lines show a moving average. d) Response function in the frequency domain, $\mathrm{Im}[\Lambda(\omega)]$, for the same parameters as in c). Note that $\mathrm{Im}[\Lambda(\omega)]$ is an odd function of $\omega$, so it suffices to plot the positive range of $\omega$. }
    \label{fig:time-traces-commutator}
\end{figure*}

\paragraph*{Application to a free-fermion model.} To illustrate the algorithm and assess its capabilities, we consider a version of the Anderson model, which describes non-interacting fermions hopping on a disordered one-dimensional lattice~\cite{anderson_absence_1958}. The Hamiltonian is given by
\begin{equation} \label{eq:aubry-andre}
    H = - J \sum_{n=1}^N \left( a_n^\dagger a_{n+1} + \mathrm{h.c.} \right)   - \lambda  \sum_{n=1}^N \epsilon_n a_n^\dagger a_n,
\end{equation}
where $a_n$ denotes the fermion annihilation operator on site $n$, $J$ is the hopping strength and $\epsilon_n$ sets the on-site potential. The parameter $\lambda$ allows us to smoothly tune the strength of the onsite potential relative to the hopping strength. We consider periodic boundary conditions such that $a_{N+1}= a_1$. This choice of a free-fermion model allows us to numerically examine our approach in large systems involving  100 sites and 20 particles using methods based on Gaussian states (see Appendix~\ref{appendix-numerics} for details).

The properties of the model depend sensitively on the choice of $\epsilon_n$. In the Anderson model, where each $\epsilon_n$ is drawn independently from a random distribution, all eigenstates are localized ~\cite{abrahams_scaling_1979}. By contrast, in the Aubry-Andr\'e (AA) model, the potential is chosen according to $\epsilon_n = \cos (2 \pi \beta n)$, where $\beta$ is an irrational number such that the potential is incommensurate with the lattice~\cite{aubry_analyticity_1980}. Translational invariance is thus broken at all length scales. In this case, the system exhibits a phase transition at $\lambda = 2J$: for $\lambda < 2J$ all states are extended, whereas for $\lambda > 2J$ all become localized. However, neither of these scenarios is desirable from our perspective as the localization properties exhibit little dependence as a function of energy. Instead, we will use a modified version of the AA model~\cite{wang_one_2020}, where
\begin{equation} \label{eq:on-site-potentail-mosaicAA}
    \epsilon_n  =  \begin{cases}
        \cos (2 \pi \beta n) \ , \  & n / \kappa \in \mathbb{Z} \ , \\
        0, \ & \mathrm{otherwise} \ . 
    \end{cases}
\end{equation}
For $\kappa = 1$, this reduces to the original AA model. Here, we will consider $\kappa = 2$ and $\beta$ equal to the golden ratio, $\beta = (\sqrt{5}+1)/2$. For a range of values of $\lambda$, extended and localized states coexist separated by mobility edges as illustrated in Fig.~\ref{fig:time-traces-commutator}(a). The mobility edges can be shown analytically to occur at energies $\pm J/\lambda$~\cite{wang_one_2020}. Throughout this work we fix $\lambda = 2$ and $J = 1$. By applying the energy filter at different energies we will hence be able to probe the qualitatively different behavior of localized and extended states.

A natural probe of the localization properties of the state is the current-current response function~\cite{scalapino_insulator_1993, takayoshi_dynamical_2022}. 

\begin{equation} \label{eq:current-current-response}
    \Lambda(\omega)  =  - i  \int_{0}^\infty dt e^{ - i \omega t}  \langle \left[ j(t), j(0) \right] \rangle,
\end{equation}
where $j = i J \sum_n (a_{n+1}^\dagger a_n - a_{n}^\dagger a_{n+1})$ is the current operator. Note that $\Lambda(\omega)$ depends on the state $\rho$ with respect to which we evaluate the expectation value $\langle \cdot \rangle = \tr{\rho \ \cdot}$, but we omit this explicit dependence for simplicity. The frequency-dependent conductivity $\sigma(\omega)$ can be expressed in terms of $\Lambda(\omega)$ as
\begin{equation} \label{eq:conductivity-frequency}
    \sigma(\omega) = i e^2 \lim_{\eta \to 0^+} \frac{\Lambda (\omega) - \langle K  \rangle}{\omega + i \eta }  ,
\end{equation}
where $e$ is the charge and $K = - J \sum_{n}(a_{n+1}^\dagger a_n + a_{n}^\dagger a_{n+1})$ the kinetic energy. For the real part of the conductivity, we obtain
\begin{equation} \label{eq:drude-regular-cond}
     \mathrm{Re}\left[\sigma(\omega) \right] =  D \delta(\omega) + \sigma^{\mathrm{reg}} (\omega)  .
\end{equation}
Here, $D =  \pi e^2( \lim_{\omega \rightarrow 0} \mathrm{Re}\left[\Lambda ( \omega) \right] - \langle K \rangle )$ is known as the Drude weight, which captures the free-particle component of the conductivity. The regular part of the conductivity is given by $ \sigma^{\mathrm{reg}} (\omega) =-e^2 \mathrm{Im}\left[\Lambda ( \omega) \right] / \omega$ . In addition to the current-current response function, we will also consider the anticommutator
\begin{equation} \label{eq:anticomm-fourier}
    \Omega (t)  = \langle  \{ \delta j(t), \delta j(0) \}  \rangle,
\end{equation}
where $\delta j(t) = j(t) - \langle j(t) \rangle$. We have again omitted the explicit state-dependence of $\Omega(t)$. This quantity captures the temporal fluctuations of the current. In thermal equilibrium, $\Lambda(\omega)$ and $\Omega(\omega)$ are related by the fluctuation-dissipation theorem~\cite{schuckert_probing_2020}. We note that the distinction between $\delta j(t)$ and $j(t)$ is unimportant for stationary states because the energy eigenstates can always be chosen to have real coefficients such that $\langle j(t) \rangle = \langle j(0) \rangle = 0$ for all eigenstates.  We will also be interested in this function in frequency domain $\Omega(\omega)$. In the computations of the Fourier transform, we use a Gaussian cutoff $e^{-t^2/2\sigma^2}$, where $\sigma$ sets the frequency resolution.

In Fig.~\ref{fig:time-traces-commutator}, we explore the current-current response function after applying energy filters of different widths $\delta$ to a Fock state $\ket{\Psi}$ with $6$ fermions distributed across $34$ sites. All fermions are initially located on odd sites such that the mean energy is $\bar{E} = 0$, corresponding to the center of the spectrum. The range of energies occupied by the initial state is captured by the local density of states (LDOS). The LDOS can be expressed in terms of the energy filter $d(E) \propto \braket{\Psi | P_{\eta}(E) | \Psi}$, where $\eta$ is a small value chosen to smoothen the discreteness of the spectrum. As shown in Fig.~\ref{fig:time-traces-commutator}(b), the local density of states takes approximately the shape of a Gaussian distribution, which is expected for a Fock state in a large system ~\cite{anshu_concentration_2016,hartmann_gaussian_2004,kuwahara_connecting_2016}. For the filtered states $P_\delta(\bar{E}) \ket{\Psi}$, the Gaussian narrows and smoothens, illustrating the ability to probe narrow ranges of energy. We highlight that computation of the LDOS of filtered states only requires measurement of the Loschmidt echo $\braket{\Psi | e^{- i H t} | \Psi}$ as a function of time $t$.

When we apply progressively narrower filters, dynamical quantities exhibit notable changes. In Fig.~\ref{fig:time-traces-commutator}(c), we plot the temporal current-current response function $\bra{\Psi} P_\delta(\bar{E}) [j(t),j(0)] P_\delta(\bar{E}) \ket{\Psi}$ for different values of $\delta$. Narrower filters enhance oscillations at low frequency and suppress oscillations at intermediate frequencies, as can be seen from the moving time-average (black line in Fig.~\ref{fig:time-traces-commutator}(c)). This information is displayed even more clearly in the frequency domain as shown in Fig.~\ref{fig:time-traces-commutator}(d): For narrow filters, the imaginary part of $\Lambda(\omega)$, which determines the regular part of the conductivity, exhibits a sharp peak close to zero frequency in addition to a cluster of peaks at high frequencies. These features can be understood from the fact that filtering effectively projects the initial state onto a delocalized eigenstate. The high-frequency peaks arise from coupling to spectrally separated localized states, whereas the low-energy peak is caused by a nearly-degenerate extended state.

For the free-fermion model considered here, the response functions depend sensitively on the choice of the initial state. Even states with similar energies may display qualitatively different response as they may be composed of disparate sets of localized and extended fermionic modes. This is to be contrasted with generic non-integrable systems, where the application of a narrow filter is expected to reproduce the microcanonical ensemble independent of the initial state~\cite{cakan_approximating_2021,lu_algorithms_2021}, as implied by, e.g., the eigenstate thermalization hypothesis~\cite{deutsch_quantum_1991, srednicki_chaos_1994,rigol_thermalization_2008}.

\begin{figure}
    \centering
    \includegraphics[width=\linewidth]{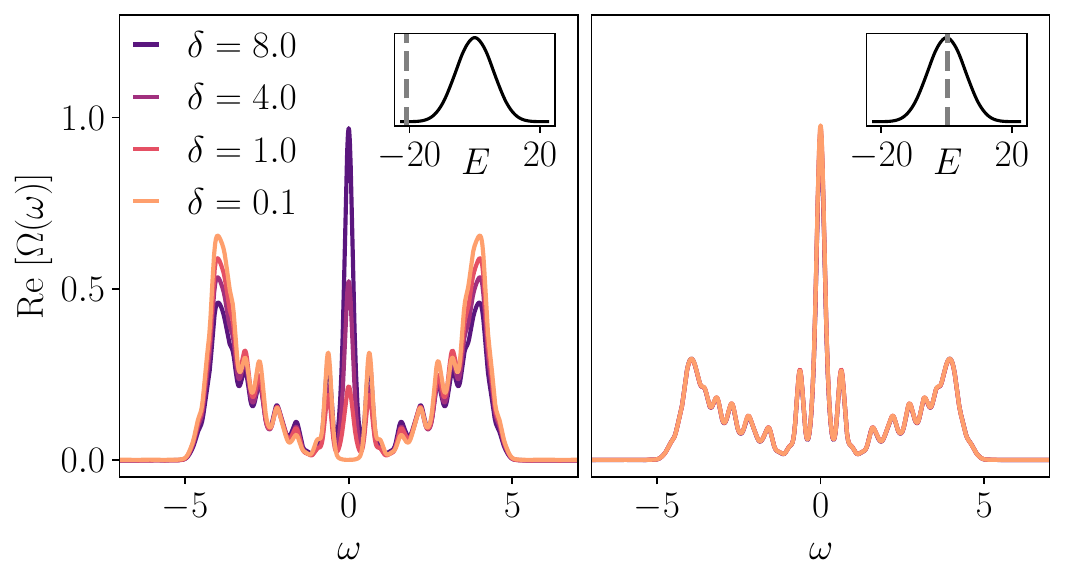}
     \caption{Filter ensemble expectation value of $\mathrm{Re} [\Omega(\omega)]$ for different filter widths $\delta$. Here $N = 34$ and $N_0 = 6$.  The inset of both figures shows the density of states (DOS), with the dashed line corresponding to the filter energy $E$. The DOS is computed through $D(E) \propto \tr{P_\nu (E)} $, with $\nu$ a small smoothing parameter. Two different energies are showcased here: The left panel shows an energy in the lower part of the spectrum, whereas the right panel displays $E = 0$ at the center.}
      \label{fig:fourier-anticomm-filter-ensemble}
\end{figure}

To circumvent this limitation of free-fermion models, we consider the filter ensemble $\rho_\delta(E) = P_\delta(E) / \tr{P_\delta(E)}$, at a given energy $E$, instead of individual initial states. For this ensemble we expect that observables vary smoothly with energy.
In Fig.~\ref{fig:fourier-anticomm-filter-ensemble} we plot the quantity $\mathrm{Re}[\Omega(\omega)]$, defined in Eq.~\eqref{eq:anticomm-fourier}, for the filter  ensemble $\rho_\delta(E)$  at two distinct energies, representing opposite regimes, and for various filter widths.  To work at a fixed filling, we have to project onto the subspace of fixed particle number $N_0$. We explain how to do this numerically in Appendix~\ref{appendix-numerics}.

The number of particles is chosen to be smaller than the number of localized modes below the mobility edge, ensuring that the multi-particle eigenstates at low energy are fully localized. In contrast, near the center of the spectrum, localized, extended, and hybrid eigenstates coexist. In the first, low-energy case (Fig.~\ref{fig:fourier-anticomm-filter-ensemble}, left), varying the filter width $\delta$ significantly changes $\mathrm{Re} [ \Omega(\omega)]$. In particular, reducing the filter width significantly suppresses the conductivity at zero frequency. This behavior arises because the eigenstates within a narrow energy window around the mean energy are localized. These states couple only weakly to other eigenstates at similar energies because the localized states are spatially well separated. Coupling to extended states incurs an energy gap and is responsible for the peaks in the conductivity at higher frequencies.  As the filter width $\delta$ increases, eigenstates with extended eigenmodes begin to contribute. These extended eigenstates can couple to other nearby extended eigenstates that are close in energy, leading to an increase in the zero-frequency conductivity. By contrast, for a mean energy near the center of the spectrum (Fig.~\ref{fig:fourier-anticomm-filter-ensemble}, right), varying the filter width has little effect. This is also expected in generic interacting systems, where the center of the spectrum corresponds to the featureless infinite temperature state.

\begin{figure}
    \centering
    \includegraphics[width=\linewidth]{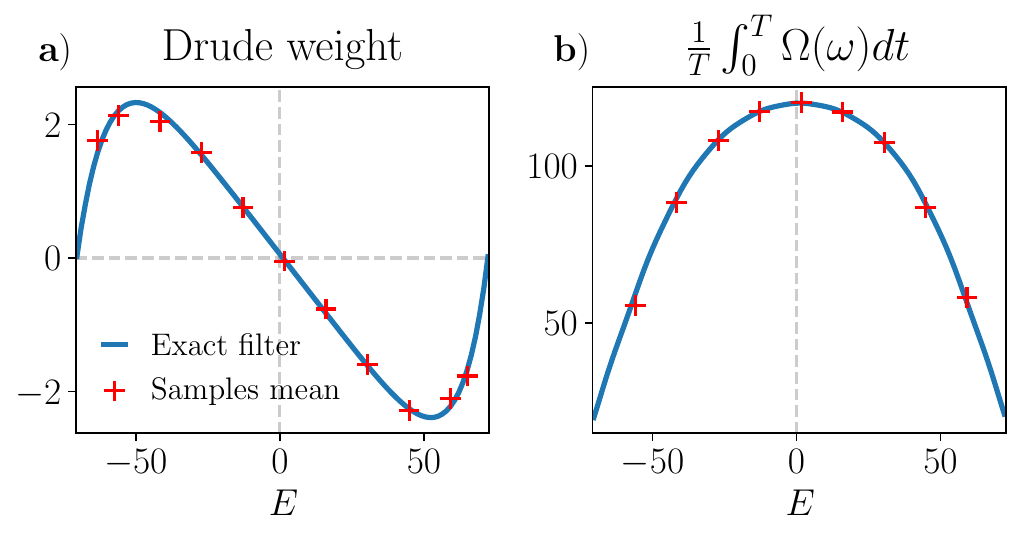}
    \caption{ a) Drude weight for the filter ensemble at different energies for $N = 100$ and filling $N_0 = 20$ with filter width $\delta = 1$. The blue line corresponds to the exact filter and the red markers to the average value of $10^4$ samples. b) Short-time average of $\frac{1}{T} \int_0^T \Omega(t) \rangle dt$ for $T = 10$. Owing to the smaller variance of this second quantity, only $10^3$ samples were used for each red marker.}
    \label{fig:filter-ensemble-sampling}
\end{figure}

Computing expectation values directly on the filter ensemble can be challenging in practice. In principle, one could evaluate them by doubling the system size and preparing a maximally entangled pair between each physical particle and an auxiliary one. However, for extensive energies $E$, the normalization factor of the filter ensemble becomes exponentially small with the system size, thus rendering this approach impractical for implementation on a quantum computer for large systems~\cite{lu_algorithms_2021}. We can overcome this issue by expressing the filter ensembles in terms of a complete set of of states $\{ \ket{\psi} \}$ as $\rho_\delta(E) \propto \sum_\psi P_{\delta/2}(E) \ket{\psi} \bra{\psi} P_{\delta/2}(E)$. As explained in Eq.~\eqref{eq:exp-value-filter-ensemble}, we can use a Markov chain to sample states from the given set according to the distribution $\bra{\psi} P_\delta(E) \ket{\psi}$ instead of iterating over all states.

Fig.~\ref{fig:filter-ensemble-sampling} illustrates the different behavior of the filter ensemble at various energies for a fixed filter width $\delta$. We show the results for both the exact filter ensemble $\rho_\delta(E)$ (blue curve), which is numerically accessible, and for an average over states sampled using Markov chain Monte Carlo (red markers). Specifically, we sample Fock states in the position basis. This set covers an extensive range of the Hamiltonian spectrum: following Eq.~\eqref{eq:on-site-potentail-mosaicAA}, a Fock state  $\ket{i_1 \cdots i_{N_0}}$ has a mean energy $\bar{E} =  \sum_{n=1}^{N_0} \epsilon_{i_n}$. We expand more on the details of the sampling procedure in Appendix~\ref{appendix-sampling}. 

In Fig.~\ref{fig:filter-ensemble-sampling}(a), we plot the Drude weight defined in  Eq.~\eqref{eq:drude-regular-cond}. The Drude weight shows an energy-dependent behavior rooted in the localization properties of the eigenstates at different energies. It vanishes for localized eigenstates, as we see at the edges of the spectrum. It also vanishes at the center of the spectrum due to the opposite signs of the effective mass at the top or bottom of the single-particle dispersion. The Drude weight is nonzero at intermediate energies, reflecting the free-particle nature of the extended eigenstates.

The crossover from localized to extended states also manifests itself in the fluctuations of the current. We quantify the low-frequency fluctuations by the time-average  $\frac{1}{T}\int_0^T \di t \, \Omega(t)$ with $T = 10$, where $\Omega(t)$ is defined in Eq.~\eqref{eq:anticomm-fourier}. Figure~\ref{fig:filter-ensemble-sampling}(b) shows that the fluctuations are largest in the middle of the spectrum as expected. For localized states, the anticommutator $\langle \{\delta j(t), \delta j(0)\} \rangle$ oscillates at high frequency around zero, resulting in a small value of the time average. For extended states, the current fluctuations remain correlated for a long time and the time average of $\Omega(t)$ is thus much larger.

\paragraph*{Outlook.}
In this work, we introduce a quantum algorithm to compute dynamical response functions of states projected onto narrow windows of energies. By probing ensembles of such states, it is thus possible to explore the dynamical response of both microcanonical and thermal expectation ensembles on a quantum computer while only probing the dynamics of an ensemble of readily preparable states. 

By performing numerical simulations of a free-fermion system, we show that the method can be used to compute the dynamical conductivity as a function of the internal energy, thereby illustrating the localization properties of the model. 

The proposed method is broadly applicable to general many-body systems, including interacting fermionic systems and spin systems. For the latter case, one may define response functions analogous to those in the text to characterize spin transport~\cite{sanchez_anomalous_2018} or consider other response functions such as the magnetic susceptibility. Quantum computers will allow one to carry out the computations presented in the text beyond classically simulable systems. We further note that the method may also be suitable for hybrid analogue-digital devices~\cite{bluvstein_quantum_2022,andersen_thermalization_2025}, where the natural time evolution of the system is combined with gates to carry out the required state preparation and measurements. 

\paragraph*{Acknowledgments.}
We thank Sirui Lu and Xiaoqi Sun for insightful discussions. This research is part of the Munich Quantum Valley, which is supported by the Bavarian State Government with funds from the Hightech Agenda Bayern Plus. This work was in part funded by the Deutsche Forschungsgemeinschaft (DFG, German Research Foundation) under Germany's Excellence Strategy – EXC-2111 – 390814868. We acknowledge funding from the German Federal Ministry of Education and Research (BMBF) through ``Efficient Quantum Algorithms for the Hubbard Model'' (EQUAHUMO) (Grant No. 13N16066), within the funding program ``Quantum Technologies—from Basic Research To Market.''


\bibliography{main}
\appendix
\onecolumngrid
\section{Numerical details}
\label{appendix-numerics}
Consider a one-dimensional lattice of $N$ sites. We work in the second quantization formalism, and $a_i^\dagger$ (resp. $a_i$) denotes the creation (resp. anihilation)  operators that create (resp. anihilate) a particle at site $i \in {1, 2, \ldots, N}$. These operators satisfy the canonical fermionic anti-commutation relations $\{ a_i, a_j^\dagger \}  = \delta_{ij}$, $\{ a_i^\dagger, a_j^\dagger \}  = \{ a_i, a_j \} = 0$. It is often useful to work in the Majorana basis, where the Majorana operators $\xi_j$ are defined as $\xi_{2j} =   {a_j+a_j^\dagger}$ and $\xi_{2j+1}   =-i (a_j-a_j^\dagger)$. They satisfy the anti-commutation relations $ \{ \xi_i, \xi_j \} = 2\delta_{ij} $.

A Hermitian operator $H$ is called quadratic if it can be expressed as
\begin{equation}
	H =  \sum_{ij} A_{ij} a_i^\dagger a_j - A^*_{ij}a_i a_j^\dagger + B_{ij}a_i a_j - B^*_{ij}a_i^\dagger a_j^\dagger \ . 
\end{equation}
The Aubry-André model presented in Eq.~\eqref{eq:aubry-andre} is an example of a quadratic Hamiltonian. 

A Gaussian state $\rho$ is defined as $\rho = e^{-H}/Z$, where  $H$ is a quadratic Hamiltonian and $Z = \tr{e^{-H}}$ a normalization factor. For such states, Wick's theorem holds: all higher-order correlation functions $\tr{\rho \xi_{i_1} \cdots \xi_{i_K}}$ factorize into combinations of two-point correlations. Consequently, the state $\rho$ is completely characterized (up to a global phase) by its covariance matrix $\Gamma$, whose elements are given by
\begin{equation}
	\Gamma_{nm} = \frac{i}{2} \mathrm{Tr} \left( \rho \left[\xi_n, \xi_m \right]\right) .
\end{equation}
The higher-order correlation functions may be expressed as $\mathrm{tr} \left[\rho \xi_{i_1} \cdots \xi_{i_K} \right] = \mathrm{Pf} \left(i \Gamma_{i_1,\ldots,i_K} \right)$, where $i_1 \leq i_2 \leq i_K \leq 2N$ and $\Gamma_{i_1,\ldots,i_K}$ denotes the submatrix of $\Gamma$ restricted to those indices~\cite{bravyi_lagrangian_2004}.


\paragraph{Filtering initial states.} Recall that the quantities of interest are given in Eqs.~\eqref{eq:loschmidt}-\eqref{eq:quantities-times}, which we reproduce here for clarity,
\begin{align} 
	f_{\psi} (t) &= \langle \psi | e^{-iH_0 t} |\psi\rangle, \\
	C^{AB}_\psi (t_1,t_2) & = \langle \psi | e^{iH_0 t_1}A   e^{-iH_0 t_2}|\psi \rangle  \ .
\end{align}
We have slightly rewritten $C^{AB}_\psi$ in terms of two times. The operator $A$ may in general depend on an additional time argument, which we omit for simplicity

The Loschmidt echo $f_{\psi} (t)$ can be equivalently expressed as $	f_{\psi} (t) =  \tr{\rho e^{-iHt}}$ with $\rho = |\Psi \rangle \langle \Psi |$. Since both $\rho$ and $e^{-iHt}$ are Gaussian operators, this trace can be evaluated using standard techniques, see, for example, Ref.~\cite{bravyi_lagrangian_2004,Bravyi_2017}.

To compute $C^{AB}_\psi (t_1,t_2)$, we rewrite it as
\begin{equation}
	\langle  \psi |  e^{iHt_1}  A e^{-iHt_2}   | \psi \rangle =  \frac{\mathrm{Tr} \left[ \rho_2 \rho_1 A  \right]}{\langle \psi_2 | \psi_1\rangle}  . 
\end{equation}
where we have denoted $\psi_{1,2} =  e^{-iHt_{1,2}} |  \psi \rangle$ and $\rho_{1,2}=  |\psi_{1,2} \rangle \langle \psi_{1,2} |$. The numerator can be computed using the identity~\cite{Bravyi_2017}
\begin{equation}
\mathrm{Tr} \left[ \rho_2 \rho_1 \xi_{i_1} \ldots   \xi_{i_K} \right] = \mathrm{Pf} \left( i \Delta^*_{i_1 \ldots i_K} \right)  \ , 
\end{equation}
where 
\begin{equation}
\Delta = (-2 \mathds{1} + i \Gamma_1 -i \Gamma_2 )(\Gamma_1 + \Gamma_2)^{-1} \ .
\end{equation}
Note that the denominator is $\langle \psi_2 | \psi_1\rangle= f_{\psi}(t_1-t_2)$, which we can compute as explained before.  
Here,  $\Gamma_{1,2}$ are the covariance matrices of $\rho_{1,2}$ and are related to the covariance matrix $\Gamma_\psi$ of the initial state $|\psi\rangle \langle \psi |$ via time evolution: 
\begin{equation}
    \Gamma_{1,2}= O(t_{1,2}) \Gamma_\psi O(t_{1,2})^T
\end{equation}
where $O(t)$ is an orthogonal matrix implementing the time evolution in the Majorana basis. In the case where $\ket{\psi}$ is a Fock state, this covariance matrix correspond to $\Gamma_\psi = \bigoplus \begin{pmatrix}
0 & \lambda_j \\ - \lambda_j & 0
\end{pmatrix}$, where $\lambda_j = 1$ if the $j$th site is empty and $\lambda_j = -1$ if it is occupied. 

\paragraph{Filter ensemble at fixed filling.} If we wish to work at a fixed filling $N_0$, we can do so by projecting onto the subspace with a fixed number of particles. This is achieved by constructing the corresponding projector
\begin{equation}
	\delta_{\hat{N},N_0} = \frac{1}{N+1} \sum_{k = 0}^{N} e^{ \frac{ i 2 \pi k (\hat{N}-N_0)}{N+1}} ,
\end{equation}
where $\hat{N} = \sum_i {a_i^\dagger a_i}$ is the particle number operator.

In particular, we will be interested in computing quantities such as $\tr{e^{-iHt} 	\delta_{\hat{N},N_0}  } $ and $\tr{e^{-iHt} 	\delta_{\hat{N},N_0} A}  $, for some operator $A$. Note that $\tr{\delta_{\hat{N},N_0}} = \begin{pmatrix}
	N_0 \\ N
\end{pmatrix}$, which is simply the dimension of the subspace at fixed filling $N_0$. In practice, the trace with the projector can be evaluated as:
\begin{equation}
	\tr{e^{-iHt} 	\delta_{\hat{N},N_0}}  = \frac{1}{N+1} \sum_{k = 0}^{N} e^{ \frac{ -i 2 \pi N_0}{N+1}}  \tr{e^{-i (Ht - \frac{2 \pi k}{N+1} \hat{N} )}} 
\end{equation}
We can define for each time step $t$ and integer $k$ a new quadratic operator $\tilde{H}(t,k) = Ht - \frac{2 \pi k}{N+1} \hat{N} $. Since $\tilde{H}(t,k)$ is quadratic, its trace can be computed using the standard Gaussian techniques discussed in the previous section.

\section{Sampling algorithm for the filter ensemble} \label{appendix-sampling} 
Here, we outline the sampling procedure for computing expectation values on the filter ensemble. From Eq.~\eqref{eq:exp-value-filter-ensemble} in the main text, we recall that
\begin{align}
   \frac{\tr{A P_\delta (E) }}{\tr{P_\delta(E)}}  & = \sum_\psi p_\psi {A}_\psi(\delta / \sqrt{2}, E) \ ,  \label{eq:exp-value-filter-ensemble-appendix}
\end{align}
where $p_\psi = \bra{\psi} P_\delta(E) \ket{\psi}$ is a probability distribution over the complete set of states $\{\psi: \sum_\psi \ket{\psi}\bra{\psi} = \mathds{1} \}$.

Then, the sum in Eq.~\eqref{eq:exp-value-filter-ensemble-appendix} can be computed by generating samples according to the probability $p_\psi$ through a Metropolis-Hasting algorithm as follows: 
\begin{itemize}
    \item Initialize the algorithm with a random state $\psi_0$. 
    \item Set up an appropriate update rule to propose a candidate $\psi_{n+1}$ conditioned on the current sample $\psi_{n}$. The update rule should be such that all configurations can be reached (ergodicity). For simplicity, we assume that the update rule is symmetric, i.e., the probability of proposing $\psi_{n+1}$ given $\psi_n$ is the same as the probability of  proposing $\psi_{n}$ given $\psi_{n+1}$.
    \item Accept the candidate $\psi_{n+1}$ with a probability $\max \{ \frac{p_{\psi_{n+1}}}{p_{\psi_{n}}} , u \}$, where $u \in [0,1]$ is a uniformly drawn random number. 
\end{itemize}
Finally, for the obtained samples $\{\psi_{1}, \ldots , \psi_{K} \}$ compute  $\frac{1}{K} \sum_i A_{\psi_i}(\delta/\sqrt{2},E)$.

For our numerical experiments, we choose the complete set of states  to be equal to the set of Fock states in position basis at fixed filling $N_0$. The initialization is done by drawing $N_0$ sites to be occupied uniformly at random. The simplest update rule is to randomly pick an occupied site and to move the fermion to a randomly chosen unoccupied site. Then, for each state $\psi_n$ we need to compute $\bra{\psi} P_\delta(E) \ket{\psi}$, which we do as explained in Appendix~\ref{appendix-numerics}. Note that only states with non-negligible LDOS $d(E)$ contribute to the sum. Since the system is integrable, we can study the convergence of the Markov Chain Monte Carlo by comparing its result with the value  of the exact filter (left-hand side of Eq.~\eqref{eq:exp-value-filter-ensemble-appendix}), computed as explained in Appendix~\ref{appendix-numerics}.
\end{document}